\documentclass[12pt]{article}
\usepackage{graphicx}
\usepackage{bbm}
\usepackage{graphics}
\usepackage{amssymb}
\usepackage{amsmath}
\usepackage[nomarginpar]{geometry}
\usepackage{setspace}
\usepackage{natbib}
\usepackage{lscape}
\usepackage{rotating}
\usepackage[tableposition=top]{caption}
\usepackage{tabularx, calc}
\usepackage{threeparttable}
\usepackage{subcaption}
\usepackage{picinpar}  
\usepackage{longtable}
\usepackage{ae}
\usepackage{float}
\usepackage{morefloats}
\usepackage{palatino}
\usepackage{hyperref}
\usepackage{color}
\usepackage{adjustbox}
\usepackage{booktabs} 
\usepackage{ulem}
\usepackage{kotex}
\usepackage{ragged2e}
\usepackage{changepage}

\definecolor{brown}{rgb}{0.5, 0.0, 0.13}
\hypersetup{
	colorlinks=true,
	citecolor=brown,
	linkcolor=red,
	pdfstartview={FitH},
	pdftitle={},
	pdfauthor={},
	pdfkeywords={}}

\begin{document}

\title{The Effect of Remote Working on Regional Waste: Evidence from South Korea\thanks{This research was supported by the Yonsei University and Yongwoon Scholarship Foundation (Yongwoon Research Grant No. 2021-11-0410).}}

\author{Hyoungchul Kim\thanks{The Wharton School, University of Pennsylvania. E-mail: \href{mailto:hchulkim@wharton.upenn.edu}{hchulkim@wharton.upenn.edu}.}  \\ \vspace{-8mm} \\ 
\small{University of Pennsylvania}
\and   Hee-Seung Yang\thanks{School of Economics, Yonsei University. E-mail: \href{mailto:heeseung.yang@yonsei.ac.kr}{heeseung.yang@yonsei.ac.kr}.}  \\ \vspace{-8mm} \\ 
\small{Yonsei University}
}

\date{March 20, 2024}
\clearpage\maketitle
\thispagestyle{empty}

\begin{abstract}
Remote working, or Work-From-Home (WFH), has become a widespread work arrangement globally. However, its broader implications for social welfare—particularly in terms of waste generation—remain underexplored. This study investigates the causal impact of remote work on regional waste production in South Korea, leveraging the COVID-19 pandemic as a natural experiment. While total household waste remains largely unchanged, regions with higher levels of WFH exhibit an increase in plastic and textile waste, alongside a decline in food waste. These findings indicate that remote working may significantly alter consumption patterns and lifestyles. Understanding these environmental effects is essential for developing sustainable work practices and effective waste management strategies in an increasingly digital and remote-oriented world.

\end{abstract}

	\clearpage\setcounter{page}{1}

\section{Introduction}

Remote working, also known as Work-From-Home (WFH), refers to work arrangement wherein employees perform their job responsibilities remotely, away from traditional office spaces. In recent years, remote working has emerged as a novel form of flexible work arrangement driven by advancements in technology (\citeauthor{bloom}, \citeyear{bloom}). In fact, remote working has experienced a notable increase in numbers. For instance, remote working rates in the U.S. grew rapidly from 0.4\% in 1965 to 4\% in 2016 (\citeauthor{barrero}, \citeyear{barrero}). This surge of remote working is not a limited trend in the U.S. The rising trend of remote working was further accelerated worldwide by the onset of the COVID-19 pandemic (\citeauthor{criscuolo}, \citeyear{criscuolo}; \citeauthor{aksoy}, \citeyear{aksoy}). Ever since the start of the COVID-19 pandemic, about 25 to 50\% of workers around the world reported that they are working remotely (\citeauthor{galasso}, \citeyear{galasso}; \citeauthor{brynjolfsson}, \citeyear{brynjolfsson}). Considering the effect of the pandemic on the job reallocation and global investments on skills and technologies supporting remote working, it is highly probable that remote working will stick as a prevalent mode of work style across various industries (\citeauthor{barrero2}, \citeyear{barrero2}).

With the increasing prominence of remote working, numerous studies have been conducted to explore the impact of this work arrangement on the labor market and productivity. In recent literature, impacts of remote working seem to be mixed. Some papers point out the multifaceted benefits of remote working: it reduces commuting time, restores work-life balance, increases autonomy for employees, and saves costs by reducing office spaces (\citeauthor{pabilonia}, \citeyear{pabilonia}; \citeauthor{harley}, \citeyear{harley}). On the other hand, some studies show that overall impacts of remote working environment were less positive and can even be detrimental to workers' productivity and performance (\citeauthor{hackney}, \citeyear{hackney}).

However, the impact of remote working on overall social welfare remains largely unexplored. This paper aims to bridge this gap by examining the influence of remote working on waste generation. Remote working can have an impact on environmental factors as it can influence lifestyle patterns. Furthermore, it can be seen as a partial process of identifying the potential effects that may arise with the proliferation of remote work. By constructing WFH measures for each district \`a la \cite{wfhmain} and using COVID-19 as an exogenous shock, this paper examines the impact of remote working on regional waste in South Korea.

Findings of the paper are manifold. The results reveal that while overall amount of waste per household is not significantly affected by the district level of remote working eligibility, remote working eligibility had significant effects  on specific types of waste. The results show that districts with higher levels of remote working eligibility experience an increase in plastic and textile waste but a decrease in food waste.

Section \ref{sec:2} explains data and construction of WFH measures used in the analysis. Sections \ref{sec:3} and \ref{sec:4} provide an empirical strategy and results of the paper. Section \ref{sec:5} shows the robustness of the main results. Section \ref{sec:6} concludes.

\section{Data}\label{sec:2}

South Korea consists of eight metropolitan cities (with Seoul being the largest) and nine provinces. These metropolitan cities and the cities within the provinces are composed of administrative subdivisions known as ``si-gun-gu.'' In the paper, we primarily focus on these si-gun-gu regions, which we will refer to as ``districts.'' Despite the presence of 229 districts in South Korea, we will consider 162 units for our analysis, as WFH measure data are aggregated for the seven metropolitan cities. Hence, a metropolitan city such as Seoul will be considered as one unit.\footnote{This has an advantage as the results will not be mainly due to variations within large metropolitan cities such as Seoul.} To summarize, we use 162 districts from years 2016 to 2021 in the analysis.\footnote{We only used year up to 2021 as our main data for the year 2022 is not yet available. We use data after 2016 to avoid potential bias in results caused by government policies concerning waste in the early 2010s.}

For the outcome variable, we use the National Waste Generation and Disposal Status (NWGDS) data from the Ministry of Environment. The NWGDS data provide annual information on household waste generation and disposal volume by district. Furthermore, as the data provide various types of waste (trash, food waste, plastic waste, etc.), we use them to check varying effects of remote working depending on the type of waste. Table \ref{tab:sum} displays the summary statistics for the NWGDS data.

\begin{table}[!htb]
\caption{Summary statistics: Average volume of waste by district (in ton)}
\centering
\begin{tabular}{lcccccccc}
\toprule
Year & Obs & All waste & Trash & Food & Plastic & Textile & Metal & Can \\
\midrule
2016 & 162 & 102426.00 & 43946.45 & 30511.30 & 5008.61 & 362.29 & 3682.66 & 1546.74 \\ 
2017 & 162 & 101408.90 & 44297.48 & 30195.41 & 5443.68 & 401.50 & 3195.55 & 1437.69 \\
2018 & 162 & 105330.20 & 45875.09 & 30101.01 & 5986.45 & 409.38 & 2767.01 & 1328.87 \\
2019 & 162 & 103462.50 & 47251.69 & 29603.18 & 5870.33 & 320.60 & 2284.09 & 891.19 \\
2020 & 162 & 106811.00 & 49641.76 & 28829.50 & 7152.84 & 508.78 & 1641.28 & 822.71 \\
2021 & 162 & 103403.20 & 50295.10 & 27626.90 & 7155.61 & 730.77 & 905.30 & 568.64 \\
\bottomrule
\multicolumn{9}{p{.95\textwidth}}{\small \textbf{Note.} Trash refers to waste that is not categorized as recyclable. As we do not use all types of waste provided by the NWGDS data due to measurement issues in certain categories, Trash, Food, Plastic, Textile, Metal, and Can waste do not sum up to the aggregate waste volume.}
\end{tabular}
\label{tab:sum}
\end{table}

In order to construct a WFH measure for each district, we need data on eligibility (in \%) for remote working by occupation category and proportion of occupations for each district. In the case of the former, we follow \citeauthor{wfhkor} (\citeyear{wfhkor}) which calculated WFH feasibility (in \%) by occupation (KSCO 1-digit)\footnote{Korea Standard Classification of Occupations, known as KSCO, is South Korea's classification of occupations which is similar to International Standard Classification of Occupations (ISCO).} in South Korea using a methodology from \citeauthor{wfhmain} (\citeyear{wfhmain}). \citeauthor{wfhmain} (\citeyear{wfhmain}) calculated the WFH feasibility ratio for each occupation in the U.S. Standard Occupational Classification (SOC 6-digit) based on 17 questions related to the eligibility of remote working from the ONET survey. To be more specific, \citeauthor{wfhmain} (\citeyear{wfhmain}) assume that occupations with a high proportion of field work, high need for equipment and device utilization, or significant direct interaction with the public during job performance are less likely to transition into remote work. Using this assumption, the eligibility for remote working is measured (assigned a value of 0 if remote working is unlikely and 1 if remote working is possible) for each specific occupation. Then the results are aggregated at the SOC 6-digit level using employment weights. \citeauthor{wfhkor} (\citeyear{wfhkor}) follows this methodology and constructs WFH feasibility by Korean Standard Occupational Classification (KSCO) 1-digit level using 2017 Korean Working Conditions Survey (KWCS). In our analysis, we follow \citeauthor{wfhmain} (\citeyear{wfhmain}) and \citeauthor{wfhkor} (\citeyear{wfhkor}) to construct WFH feasibility by 2-digit level to gain a finer measure for WFH eligibility.\footnote{Note that our results are similar even if we use a rougher 1-digit level.} We provide the table of WFH eligibility by occupation (KSCO 1- and 2-digit) in the appendix. 

For data on proportion of occupations in each district, we use the Local Area Labour Force Survey (LALFS). The LALFS is a biannual survey conducted by the Ministry of Statistics in South Korea. The survey collects data from a representative sample of approximately 200,000 households, focusing on household members who are 15 years of age or older. Each wave of the survey involves approximately 400,000 respondents. The primary focus of the survey is the economic activity of the participants, specifically variables related to their employment status. Additionally, it includes socio-demographic information of the respondents, such as age, gender, education level, income, and place of residence. As the survey provides an individual's occupation type based on KSCO, we use it to calculate the proportion of occupation categories for each district.

Combining these two data, we construct a WFH measure for each district ($\mathrm{WFH}_i$).\footnote{\citeauthor{wfhmain} (\citeyear{wfhmain}) and \citeauthor{wfhkor} (\citeyear{wfhkor}) also provide a WFH measure by industry classification. Even though we do not use it as our main explanatory variable, we use it as a control variable to account for the effect of industry composition.} The construction of the WFH measure is given as follows: 

\begin{align}
    \mathrm{WFH}_i=\sum_j \frac{\mathrm{Work}_{ij}}{\mathrm{Work}_i} \cdot \mathrm{WFH}_j,
\end{align}

\noindent where $i$ and $j$ denote district and occupation, respectively. $\mathrm{WFH}_j$ is the WFH measure of occupation $j$. $\mathrm{Work}_i$ denotes the total number of workers in district $i$ and $\mathrm{Work}_{ij}$ denotes the total number of workers in occupation $j$ in district $i$. By construction, the measure ranges from 0 to 1. In the analysis, we use the WFH measure for the year 2019, which is just before the onset of COVID-19 in 2020. However, the baseline results remain similar even if we use the measure from other available years.

\begin{figure}[!htb]
	\begin{center}
		\includegraphics[width=1.05\textwidth]{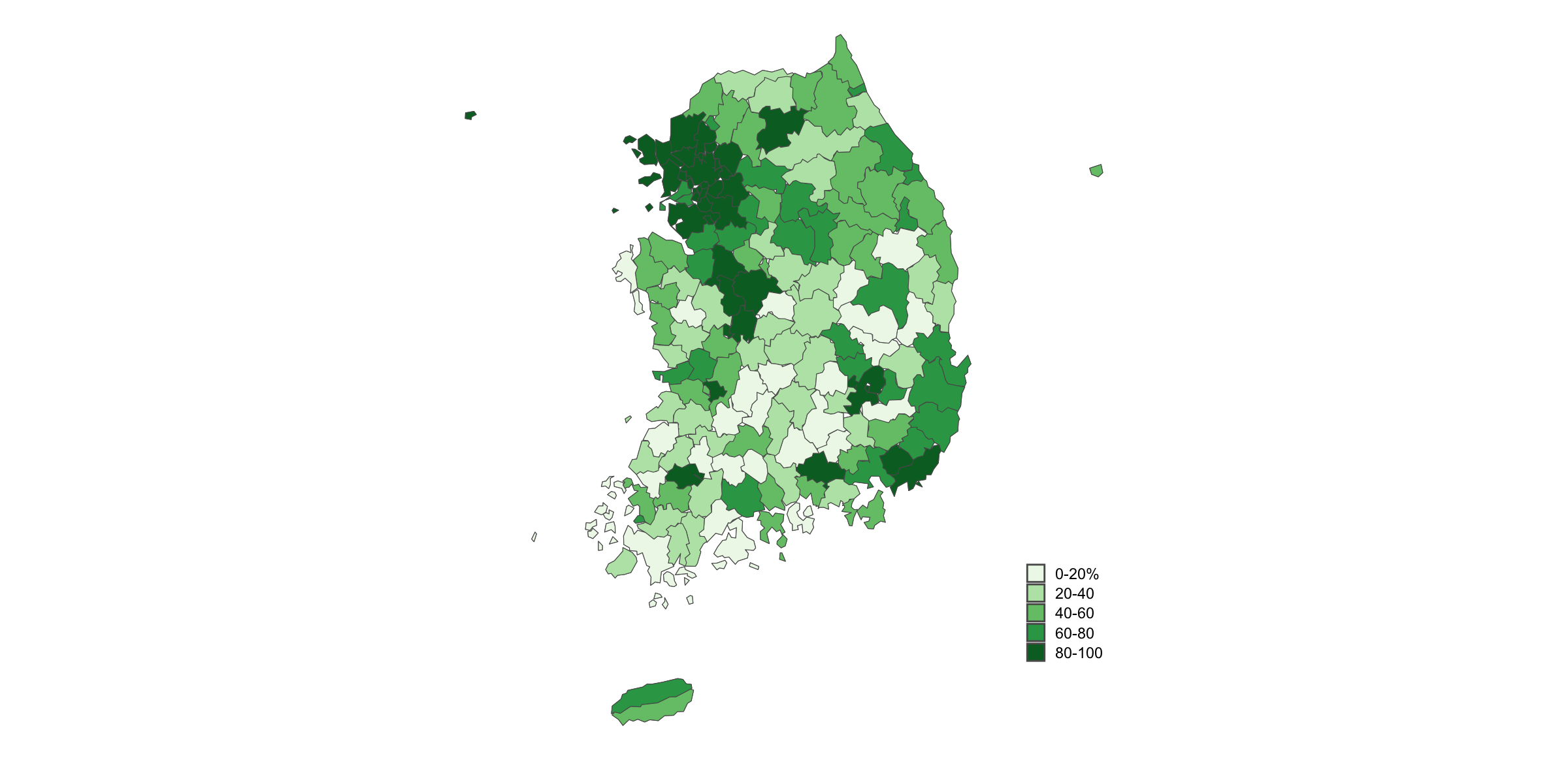} 
	\end{center}
        \caption{WFH measure by district (2019)}\label{fig:wfh}
\end{figure}

Figure \ref{fig:wfh} plots the distribution of the WFH measure (by quantile) in 2019 by district in South Korea. Figure \ref{fig:wfh} illustrates that the WFH measure is not spatially uniform. The uneven distribution demonstrates that remote working is more pervasive in certain areas. We also provide distribution of the WFH measure across district by year in the appendix. We can easily see in Figure \ref{fig:wfh2016} that the distribution of WFH measure is fairly consistent over the years.

We also use data from the Ministry of Statistics and Ministry of the Interior and Safety to secure various control variables (mean age, proportion of population under the age of 30, initial population in 2016, etc. ) for each district. 

Finally, we conducted a survey ``Yonsei-Yongwoon Daily Time Use Survey'' in July 2022 in Korea, involving 3,000 individuals (representative sample of South Korea) with the assistance of a survey company. This survey provides us with insights into how people manage their time, taking into account different family types or work statuses. To be more specific, the survey provides information on how people allocate their time across various daily activities, such as housework, work, sleep, personal study, and more. The strength of this survey is that it also includes a variable on whether people work from home. Using this variable and people's responses on daily time use, we evaluate if the characteristics of remote workers in South Korea are aligned with the general characteristics of remote workers documented in previous literature. Furthermore, we employ this survey to investigate potential underlying factors that might have driven our primary findings in Section \ref{sec:4}.

\section{Empirical strategy}\label{sec:3}

In order to estimate the causal effect of remote working on regional waste, we employ the following Difference-in-Differences (DiD) equation: 
\begin{align}
    \frac{y_{it}}{H_{i,2016}}=\alpha+\beta\mathrm{Post}_t \times \mathrm{WFH}_i+\gamma X_{it}'+\delta_d+\delta_t+\delta_d Y_t +\varepsilon_{it},\label{eq:did}
\end{align}

\noindent where the dependent variable $\frac{y_{it}}{H_{i,2016}}$ denotes waste per household. $y_{it}$ corresponds to the amount of waste (by ton) and $H_{i,2016}$ is the total number of households in the initial period (2016) in district $i$. Standardizing the dependent variable by the initial number of households removes the inherent specification bias caused by scale effects (\citeauthor{peri}, \citeyear{peri}).\footnote{Note that dividing $y_{it}$ by initial population instead of households gives similar results. Also, dividing $y_{it}$ by yearly population or yearly households gives similar results. This is elaborated in Section \ref{sec:5}.} $\mathrm{Post}_t$ is a dummy variable that captures the onset of the exogenous COVID-19 shock in 2020. It becomes one when the year $t$ is 2020 and 2021, and zero otherwise. $\mathrm{WFH}_i$ is the WFH measure that accounts for the district $i$'s proportion of workers eligible for remote working. $X_{it}$ indicates region specific control variables such as initial population, mean age, district's industrial characteristics, population proportion of people in their 30s or younger, average income, GRDP, and sex ratio. Lastly, $\delta_d Y_t$ denotes location-specific time trends while $\delta_t$ and $\delta_d$ denote year fixed effect and location fixed effects, respectively. As we are comparing changes between districts in South Korea, our results can be biased if there are some varying trends across regions. Hence we add location-specific time trends to account for these heterogeneous trends that could affect our results. For location-specific time trends and location fixed effects, we employ Travel-to-Working Areas given by \citeauthor{cz} (\citeyear{cz}). This is similar to the idea of commuting zone in the U.S. Districts in South Korea are grouped into 34 regions.

The coefficient of interest is $\beta$ as it captures the causal effect of the level of the WFH measure on household waste, using COVID-19 as an exogenous shock. Throughout the paper, the results will show the estimation results of the coefficient.

\section{Results}\label{sec:4}

Table \ref{tab:did2} shows the estimates of the DiD coefficient in Equation \ref{eq:did}. The outcome variables are various types of waste per household. The dependent variables include all waste, trash, food, plastic, textile, metal and can waste (in ton) per household. Table \ref{tab:did2} shows that variations in the WFH measure did not significantly affect the amount of aggregate waste per household. This suggests that the proportion of remote working in a district may not have a statistically significant effect on the amount of aggregate waste per household. 

However, Table \ref{tab:did2} shows that remote working significantly affected the amount of food, plastic and textile waste. Notably, a higher level of WFH measure in a district causes an increase in plastic and textile waste. For example, a one-unit increase in the WFH measure raises plastic and textile waste by 0.031 and 0.01 tons per households, respectively. This imply that remote working environment may contribute to increased consumption and disposal of plastic and textile products. This finding suggests that WFH arrangements may lead to changes in consumer behavior and lifestyle, potentially resulting in increased waste generation in these categories.

\begin{table}[!htb]
\caption{Main results}
\centering
\small % Reduce font size
\setlength{\tabcolsep}{4pt} % Adjust column spacing
\begin{tabular}{l *{7}{c}}
\toprule
& $(1)$ & $(2)$ & $(3)$ & $(4)$ & $(5)$ & $(6)$ & $(7)$ \\
\multicolumn{1}{c}{Dependent Variable} & All Waste & Trash & Food & Plastic & Textile & Metal & Can \\
\midrule
$\mathrm{Post} \times \mathrm{WFH}$ & $-0.039$ & $0.061$ & $-0.110^{**}$ & $0.031^{**}$ & $0.010^{***}$ & $-0.006$ & $-0.010$\\
& $(0.128)$ & $(0.101)$ & $(0.051)$ & $(0.014)$ & $(0.004)$ & $(0.025)$ & $(0.011)$\vspace{1mm} \\
Adjusted $R^2$ & $0.481$ & $0.434$ & $0.475$ & $0.340$ & $0.141$ & $0.307$ & $0.256$\\
$\mathrm{N}$ & $972$ & $972$ & $972$ & $972$ & $972$ & $972$ & $972$\\
\bottomrule
\multicolumn{8}{p{.9\textwidth}}{\small \textbf{Note.} Standard errors in parentheses are clustered at the district ($N=162$) level. $^{***}~\, \mathrm{p}<~0.01$; $^{**}\, \mathrm{p}< 0.05$ ; $^{*} \, \mathrm{p}<0.10$.}
\end{tabular}
\label{tab:did2}
\end{table}

Interestingly, while districts with a higher level of WFH measure experienced an increase in plastic and textile waste, food waste decreased in those areas. This could be attributed to reduced food consumption outside the home, as individuals have more flexibility to prepare meals at home while working remotely (\citeauthor{restrepo}, \citeyear{restrepo}). Thus, the decline in food waste may indicate a positive environmental outcome associated with WFH arrangements. Also, previous literature points out that remote working leads to heavier workload and lunch disrupt (\citeauthor{wu}, \citeyear{wu}, \citeauthor{skynova}, \citeyear{skynova}). Hence, reduced food waste may be due to the disruption of food consumption in the remote working environment. 

\begin{figure}
    \begin{center}
    \includegraphics[width=\textwidth]{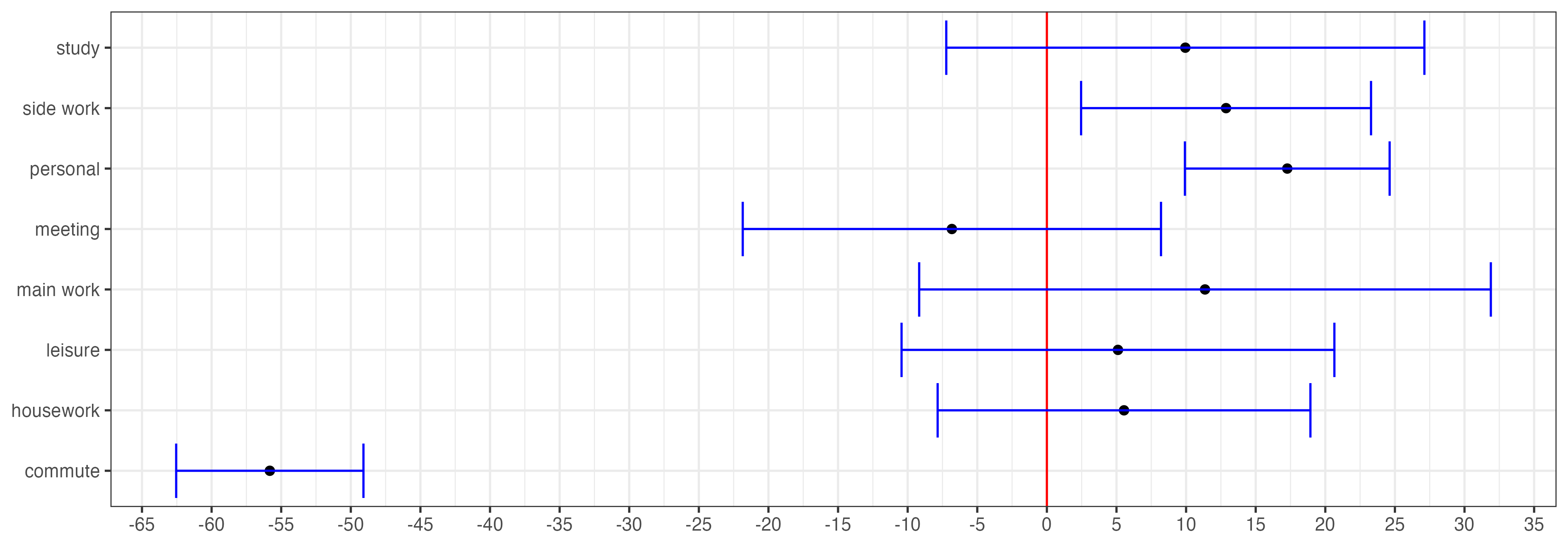}
    \caption{Effect of WFH on Daily Time Management (per minute)}
    \label{fig:timeuse}
    \end{center}
    \footnotesize \textbf{Note:} This figure plots the estimated coefficient on the effect of WFH on daily time use. The 90 percent confidence intervals are calculated with standard errors clustered at the regional (Si-do) level.
\end{figure}

We further use ``Yonsei-Yongwoon Daily Time Use Survey'' to support our conjectures. Figure \ref{fig:timeuse} shows the t-test results comparing WFH individuals and non-WFH individuals and plot the coefficients with 90 percent confidence intervals. The coefficient bars indicate the effect of WFH on daily time use patterns. First, the figure shows that daily time use patterns of remote workers in South Korea seem to be aligned with the general characteristics of remote workers in past literature (\citeauthor{pabilonia}, \citeyear{pabilonia}; \citeauthor{harley}, \citeyear{harley}). For instance, the data indicate that remote workers allocate more time to personal care activities. Particularly, there is a great reduction of time spent on commute. These patterns imply that remote workers in South Korea follow similar patterns mentioned in past literature. This suggests that remote working increases autonomy and personal time for employees due to the flexibility in the work arrangement. Considering the fact that increased activity in home leads to greater home consumption, this could be a contributing factor to the rise in plastic waste generation among remote workers. Our data also reveal that on average remote workers tend to do more side-work than non-remote workers. This could indicate that the remote working arrangement is leading to additional workload to the workers, as reported in previous literature. Thus, this could possibly explain the significantly lower level of food waste in our main analysis.

Still, our DiD regression results might be due to some pre-existing trends before the treatment period. Hence, we also plot event study coefficients including leads and lags of the treatment to test whether the parallel trend assumption hold.\footnote{Note that we only plot the outcomes that are statistically significant (food, plastic, and textile waste).} The equation for plotting the event study is as follows: 
\begin{align}
    \frac{y_{it}}{H_{i,2016}}=\alpha+ \sum_{r \neq 2019} \left( \beta_r 1\{ r = t \} \times 
 \mathrm{WFH}_i \right)+\gamma X_{it}'+\delta_d+\delta_t+\delta_d Y_t +\varepsilon_{it}.\label{eq:trend}
\end{align}

\begin{figure}[!tb]
  \centering
  \begin{subfigure}{0.32\textwidth}
    \centering
    \includegraphics[width=\linewidth]{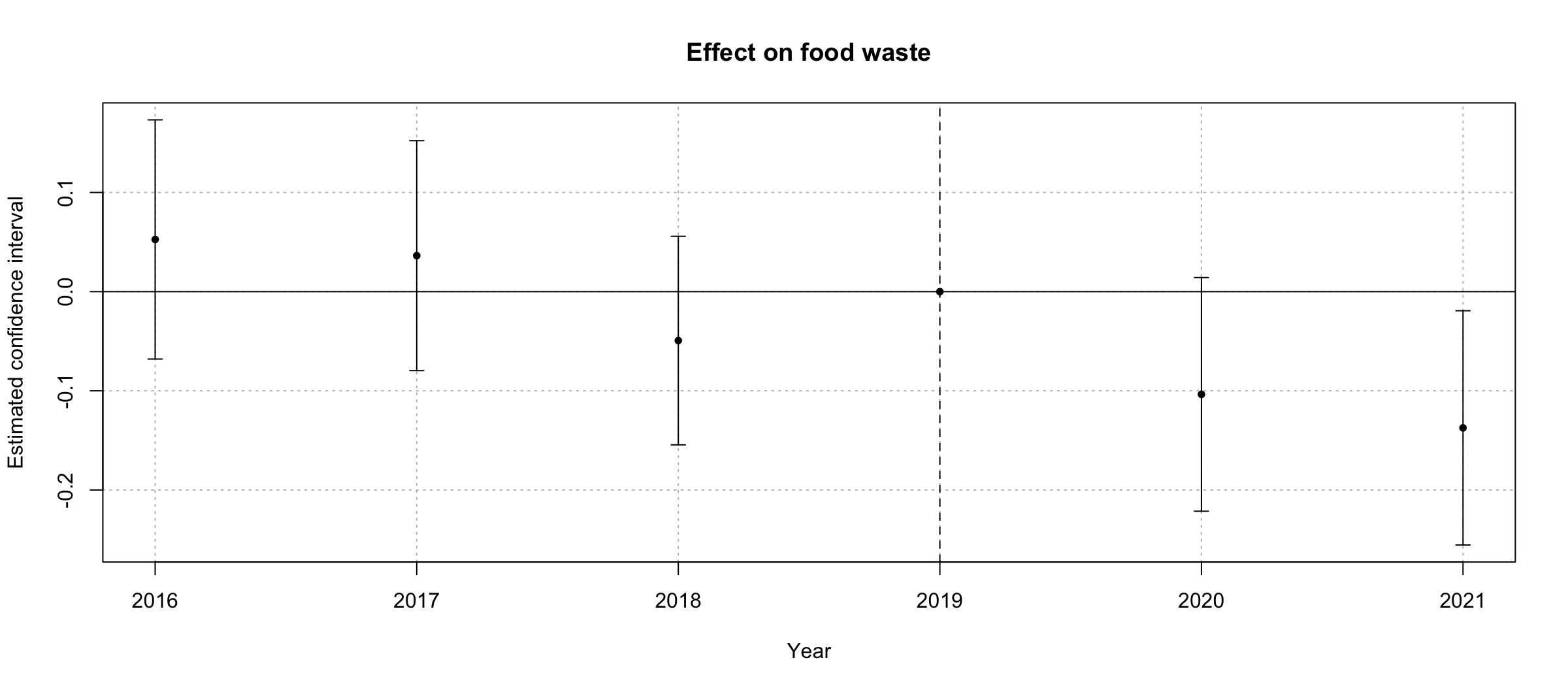}
    \label{fig:food}
  \end{subfigure}
  \begin{subfigure}{0.32\textwidth}
    \centering
    \includegraphics[width=\linewidth]{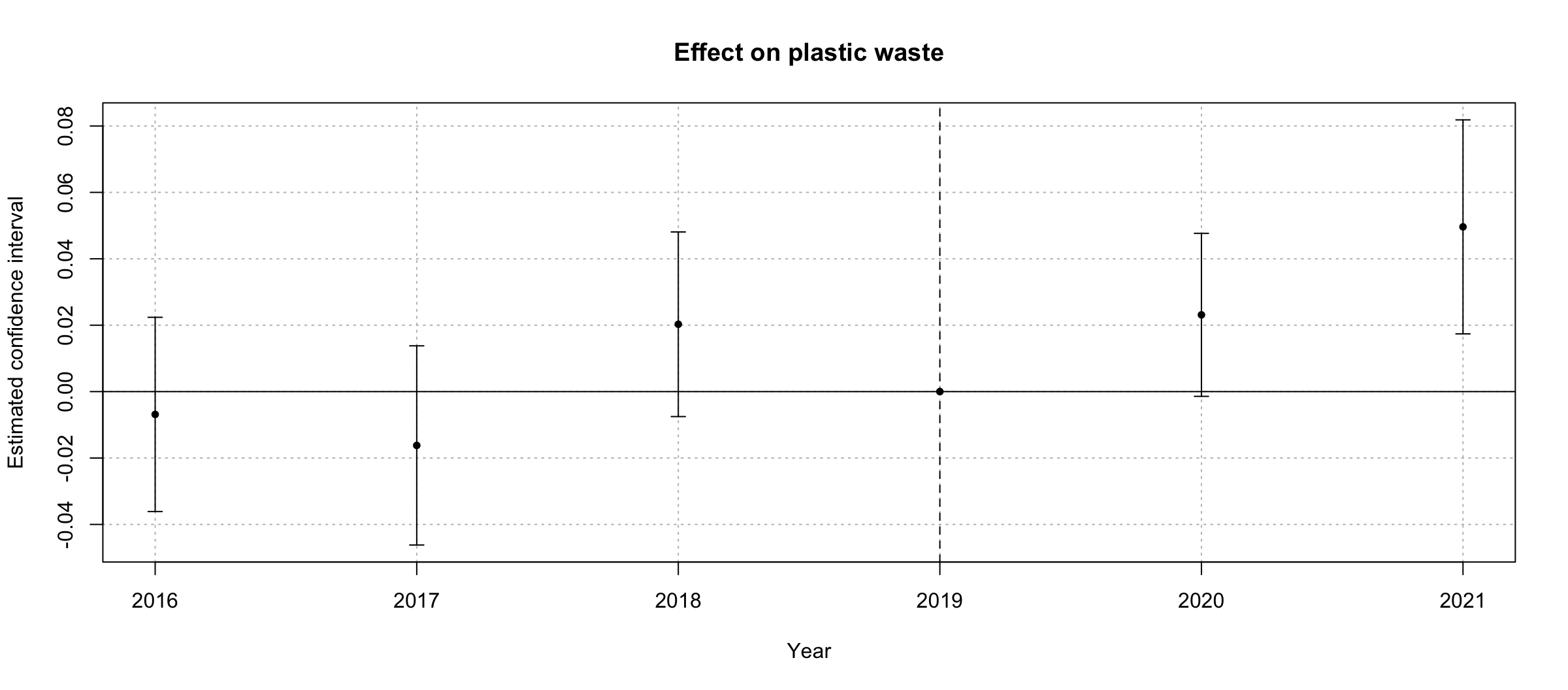}
    \label{fig:plastic}
  \end{subfigure}
  \begin{subfigure}{0.32\textwidth}
    \centering
    \includegraphics[width=\linewidth]{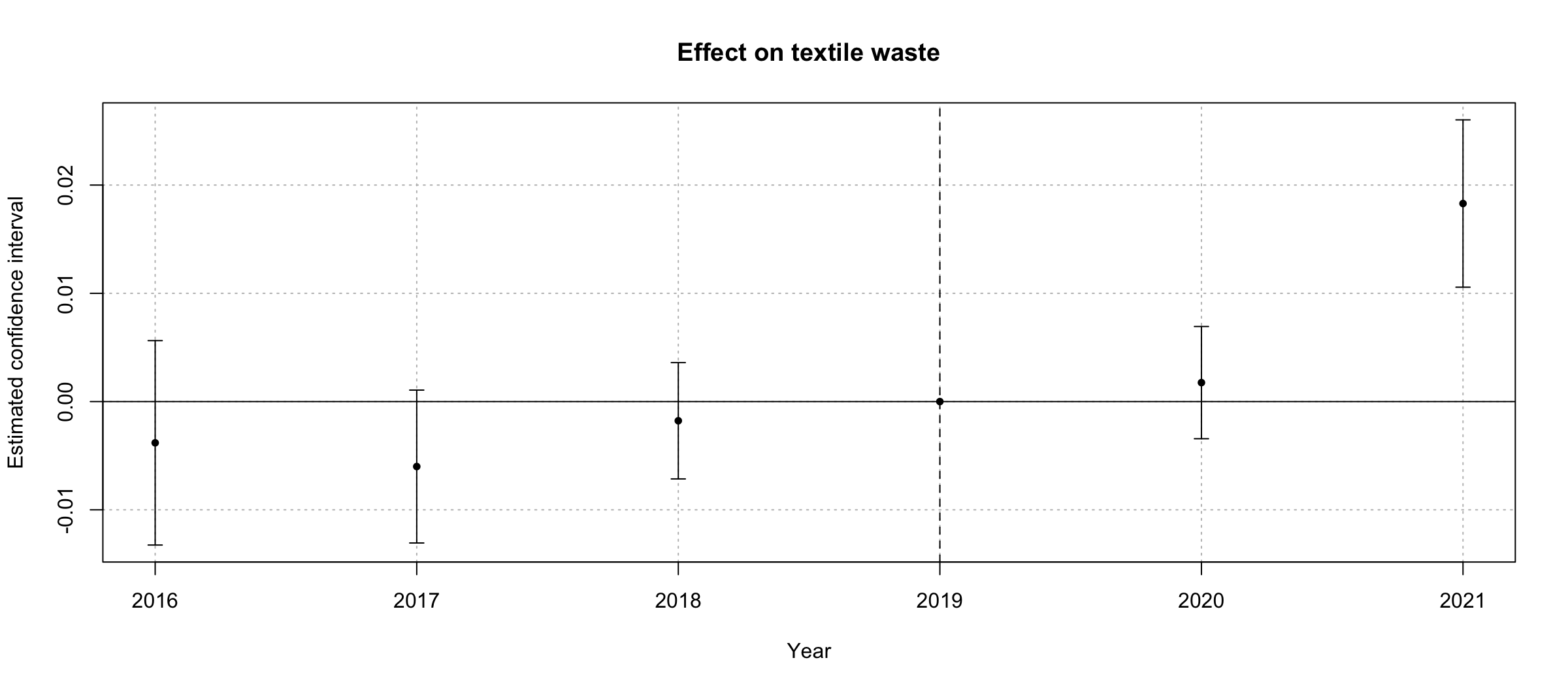}
    \label{fig:textile}
  \end{subfigure}
  \caption{Estimated 90\% confidence interval for DiD coefficients}
  \label{fig:trend}
\end{figure}

Figure \ref{fig:trend} plots the DiD coefficient and its 90 percent confidence intervals. As shown in the figure, the point estimates are not statistically significant until after 2020. This pattern aligns with the parallel trends assumption that underlies our DiD analysis, providing additional support for the empirical findings of the paper.

\section{Robustness checks}\label{sec:5}

As our data are aggregated on an annual basis, it is unlikely that the results of the paper are severely affected by the regional variations of the COVID-19 intensity. The main results indirectly control for this possibility by adding initial district population as a control variable since the number of COVID-19 patients are proportional to the district population. To further control for possible regional differences, we re-estimate the analysis by excluding Seoul, Daegu, and Gyeongsangbuk-do. This is because there was a noticeable surge of COVID-19 cases in these regions during 2020. Hence, the main results could be biased if our results are mainly due to the effect in these areas. Table \ref{tab:did_noseoultk} shows result without these regions. Table \ref{tab:did_noseoultk} clearly shows that the results are similar to the main regression results. Thus, the main results of this paper is not driven by possible regional differences. Additionally, we add the share of total number of COVID-19 patients (year 2020) by initial period population in each province to our baseline estimation to further control for the possibility of regional differences due to COVID-19 intensity.\footnote{Due to data limit, we do not have a finer unit of data for the number of COVID-19 patients than the nine province level. However, our robustness checks using the number of COVID-19 patients in each province and excluding COVID-19 intensive regions show credible evidence that our results are not mainly driven by regional differences. As COVID-19 soon became a common shock across all regions in South Korea, it is likely that the regional difference due to COVID-19 intensity does not vary greatly between small administrative units. We utilized data from COVID-19 patients throughout the year 2020, primarily due to the presence of regional fluctuations in the number of COVID-19 cases up until June 2020.} Table \ref{tab:covid-control} shows that our results are not particularly sensitive to the regional difference of COVID-19 intensity.

\begin{table}[!htb]
\caption{Regression results with excluded regions (Seoul, Daegu, Gyeongsangbuk-do)}
\centering
\begin{tabular}{lcccc}
\toprule & $(1)$ & $(2)$ & $(3)$ & $(4)$ \\
\multicolumn{1}{c}{ Dependent Variable}  & All Waste & Food & Plastic & Textile \\
\midrule
$\mathrm{Post} \times \mathrm{WFH} $ & $-0.046$  & $-0.111^{*}$ & $0.033^{**}$ &$0.011^{***}$ \\ 
& $(0.141)$ & $(0.058)$ & $(0.016)$ & $(0.004)$\vspace{1mm} \\ 
Adjusted $R^2$ & $0.517$ & $0.487$ & $0.324$ & $0.134$\\
$\mathrm{~N}$ & $822$ & $822$ & $822$ & $822$\\
\bottomrule
\multicolumn{5}{p{.8\textwidth}}{\small \textbf{Note.} Standard errors in parentheses are clustered at the district ($N=137$) level. $^{* * *} \, \mathrm{p}<0.01$; $^{**}\, \mathrm{p}< 0.05$ ; $^{*} \, \mathrm{p}<0.10$.}
\end{tabular}\label{tab:did_noseoultk}
\end{table}

\begin{table}[!htb]
\caption{Regression results with number of COVID-19 patients}
\centering
\begin{tabular}{lcccc}
\toprule & $(1)$ & $(2)$ & $(3)$ & $(4)$ \\
\multicolumn{1}{c}{ Dependent Variable}  & All Waste & Food & Plastic & Textile \\
\midrule
$\mathrm{Post} \times \mathrm{WFH} $ & $-0.054$  & $-0.113^{**}$ & $0.033^{**}$ &$0.011^{***}$ \\ 
& $(0.132)$ & $(0.051)$ & $(0.014)$ & $(0.004)$\vspace{1mm} \\ 
Adjusted $R^2$ & $0.483$ & $0.475$ & $0.345$ & $0.140$\\
$\mathrm{~N}$ & $972$ & $972$ & $972$ & $972$\\
\bottomrule
\multicolumn{5}{p{.8\textwidth}}{\small \textbf{Note.} Standard errors in parentheses are clustered at the district ($N=162$) level. $^{* * *} \, \mathrm{p}<0.01$; $^{**}\, \mathrm{p}< 0.05$ ; $^{*} \, \mathrm{p}<0.10$.}
\end{tabular}\label{tab:covid-control}
\end{table}

In order to show that our baseline results are not sensitive to the change in the construction of the WFH measure, we also rerun our estimation employing a rougher measure with 1-digit KSCO occupation level. Table \ref{tab:did1} shows that the results are similar to the main regression results. Thus the main results of this paper are not particularly sensitive to the construction of the WFH measure.

\begin{table}[!htb]
\caption{Regression results using KSCO occupation 1-digit WFH measure}
\centering
\small % Reduce font size
\setlength{\tabcolsep}{4pt} % Adjust column spacing
\begin{tabular}{l *{7}{c}}
\toprule
& $(1)$ & $(2)$ & $(3)$ & $(4)$ & $(5)$ & $(6)$ & $(7)$ \\
\multicolumn{1}{c}{Dependent Variable} & All Waste & Trash & Food & Plastic & Textile & Metal & Can \\
\midrule
$\mathrm{Post} \times \mathrm{WFH}$ & $-0.062$ & $0.040$ & $-0.112^{**}$ & $0.035^{**}$ & $0.011^{***}$ & $-0.009$ & $-0.011$\\
& $(0.129)$ & $(0.101)$ & $(0.050)$ & $(0.014)$ & $(0.004)$ & $(0.025)$ & $(0.011)$\vspace{1mm} \\
Adjusted $R^2$ & $0.481$ & $0.434$ & $0.475$ & $0.340$ & $0.141$ & $0.307$ & $0.256$\\
$\mathrm{N}$ & $972$ & $972$ & $972$ & $972$ & $972$ & $972$ & $972$\\
\bottomrule
\multicolumn{8}{p{.9\textwidth}}{\small \textbf{Note.} Standard errors in parentheses are clustered at the district ($N=162$) level. $^{***}~\, \mathrm{p}<~0.01$; $^{**}\, \mathrm{p}< 0.05$ ; $^{*} \, \mathrm{p}<0.10$.}
\end{tabular}
\label{tab:did1}
\end{table}

As the results can be affected by possible spillovers across local labor markets, we re-estimate the result using 17 region fixed effect (including 8 metropolitan cities and nine provinces). Table \ref{tab:did_sido} clearly shows that the results are similar to the main regression specification, further supporting the main results.

\begin{table}[!htb]
\caption{Regression results with 17 region fixed effect and linear trends}
\centering
\begin{tabular}{lcccc}
\toprule & $(1)$ & $(2)$ & $(3)$ & $(4)$ \\
\multicolumn{1}{c}{ Dependent Variable}  & All Waste & Food & Plastic & Textile \\
\midrule
$\mathrm{Post} \times \mathrm{WFH} $ & $-0.062$  & $-0.115^{**}$ & $0.035^{**}$ &$0.012^{***}$ \\ 
& $(0.143)$ & $(0.052)$ & $(0.015)$ & $(0.004)$\vspace{1mm} \\ 
Adjusted $R^2$ & $0.398$ & $0.414$ & $0.326$ & $0.080$\\
$\mathrm{~N}$ & $972$ & $972$ & $972$ & $972$\\
\bottomrule
\multicolumn{5}{p{.8\textwidth}}{\small \textbf{Note.} Standard errors in parentheses are clustered at the district ($N=162$) level. $^{* * *} \, \mathrm{p}<0.01$; $^{**}\, \mathrm{p}< 0.05$ ; $^{*} \, \mathrm{p}<0.10$.}
\end{tabular}\label{tab:did_sido}
\end{table}

We also re-estimate equation \ref{eq:did} by dividing amount of waste by the initial number of population in the district in the dependent variable. This is to show that the main results in the paper are not sensitive to the change in the method of standardization. Table \ref{tab:did_pop} demonstrates that the impact of the WFH measure on regional waste remains qualitatively consistent with the main findings, when the amount of waste is standardized by district population in the initial year.
 
\begin{table}[!htb]
\caption{Regression results using dependent variables standardized by initial population}
\centering
\begin{tabular}{lcccc}
\toprule & $(1)$ & $(2)$ & $(3)$ & $(4)$ \\
\multicolumn{1}{c}{ Dependent Variable}  & All Waste & Food & Plastic & Textile \\
\midrule
$\mathrm{Post} \times \mathrm{WFH} $ & $-0.010$  & $-0.044^{**}$ & $0.012^{**}$ &$0.004^{***}$ \\ 
& $(0.057)$ & $(0.021)$ & $(0.006)$ & $(0.002)$\vspace{1mm} \\ 
Adjusted $R^2$ & $0.475$ & $0.374$ & $0.284$ & $0.131$\\
$\mathrm{~N}$ & $972$ & $972$ & $972$ & $972$\\
\bottomrule
\multicolumn{5}{p{.8\textwidth}}{\small \textbf{Note.} Standard errors in parentheses are clustered at the district ($N=162$) level. $^{* * *} \, \mathrm{p}<0.01$; $^{**}\, \mathrm{p}< 0.05$ ; $^{*} \, \mathrm{p}<0.10$.}
\end{tabular}\label{tab:did_pop}
\end{table}

\section{Conclusion}\label{sec:6}

This paper examines the impact of remote working on regional waste in South Korea. Even though remote working has already become one of the prevalent working arrangement, its impact on the overall social welfare is still largely unexplored. This paper fills this gap by analyzing the causal impact of remote working on waste generation at the region level. Using COVID-19 as an exogenous shock for the regional variation in the WFH measure, we employ a DiD approach to examine the environmental implications of remote working. Our findings show that the overall amount of waste per household is not significantly affected by the regional variation in the WFH measure. However, there are significant effects on specific types of waste. The districts with a higher level of WFH measure experience an increase in plastic and textile waste but a decrease in food waste. 

Overall, this paper sheds lights on the effect of remote working environment on regional waste generation. The results contribute to the growing body of literature on remote working and its implications for social welfare. Understanding the environmental consequences of remote working is critical as it is likely to stick as one of the prevalent mode of work style. In that sense, the results of the paper is crucial for policymakers, organizations, and individuals aiming to develop sustainable practices and understand the environmental impact of remote working.

\newpage
 \bibliographystyle{aer}
\bibliography{won}

@article{wfhkor,
author={Sungwoong Choi},
title={Characteristics and Distribution of Teleworkable Jobs Based on Physical Working Conditions (in Korean)},
journal={Journal of the Economic Geographical Society of Korea},
issn={1226-8968},
year={2020},
volume={23},
number={3},
pages={276-291},
doi={10.23841/egsk.2020.23.3.276}
}

@article{wfhmain,
title = {How many jobs can be done at home?},
journal = {Journal of Public Economics},
volume = {189},
pages = {104235},
year = {2020},
issn = {0047-2727},
doi = {https://doi.org/10.1016/j.jpubeco.2020.104235},
url = {https://www.sciencedirect.com/science/article/pii/S0047272720300992},
author = {Jonathan I. Dingel and Brent Neiman},
}

@article{cz,
author={Sewon Lee and Hee-Yeon Lee},
title={Delimitation of City-Regions Based on the Method of Travel-to-Working Area and Analyzing Spatial Structure of City-Regions (in Korean)},
journal={The Korea Spatial Planning Review},
issn={1229-8638},
year={2015},
volume={84},
pages={165-189},
doi={10.15793/kspr.2015.84..010}
}

@article{peri,
title = {Assessing inherent model bias: An application to native displacement in response to immigration},
journal = {Journal of Urban Economics},
volume = {69},
number = {1},
pages = {82-91},
year = {2011},
issn = {0094-1190},
doi = {https://doi.org/10.1016/j.jue.2010.08.005},
url = {https://www.sciencedirect.com/science/article/pii/S0094119010000628},
author = {Giovanni Peri and Chad Sparber}
}

@TechReport{barrero,
  title={The Evolution of Working from Home},
  author={Barrero, Jose Maria and Bloom, Nicholas and Davis, Steven J.},
  year={2023},
  institution={Institute for Economic Policy Research}
}

@techreport{barrero2,
 title = {Why Working from Home Will Stick},
 author = {Barrero, Jose Maria and Bloom, Nicholas and Davis, Steven J},
 institution = {National Bureau of Economic Research},
 type = {Working Paper},
 series = {Working Paper Series},
 number = {28731},
 year = {2021},
 month = {April},
 doi = {10.3386/w28731},
 URL = {http://www.nber.org/papers/w28731},
}

@techreport{galasso,
   author = {Vincenzo Galasso and Martial Foucault},
   title = {Working during COVID-19},
   year = {2020},
   number = {246}, 
   url = {https://www.oecd-ilibrary.org/content/paper/34a2c306-en},
   doi = {https://doi.org/https://doi.org/10.1787/34a2c306-en}
}

@techreport{brynjolfsson,
 title = {COVID-19 and Remote Work: An Early Look at US Data},
 author = {Brynjolfsson, Erik and Horton, John J and Ozimek, Adam and Rock, Daniel and Sharma, Garima and TuYe, Hong-Yi},
 institution = {National Bureau of Economic Research},
 type = {Working Paper},
 series = {Working Paper Series},
 number = {27344},
 year = {2020},
 month = {June},
 doi = {10.3386/w27344},
 URL = {http://www.nber.org/papers/w27344},
}

@techreport{criscuolo,
   author = {Chiara Criscuolo and Peter Gal and Timo Leidecker and Francesco Losma and Giuseppe Nicoletti},
   title = {The role of telework for productivity during and post-COVID-19},
   year = {2021},
   number = {31}, 
   url = {https://www.oecd-ilibrary.org/content/paper/7fe47de2-en},
   doi = {https://doi.org/https://doi.org/10.1787/7fe47de2-en}
}

@Article{pabilonia,
  author={Sabrina Wulff Pabilonia and Victoria Vernon},
  title={{Telework, Wages, and Time Use in the United States}},
  journal={Review of Economics of the Household},
  year=2022,
  volume={20},
  number={3},
  pages={687-734},
  month={September},
  doi={10.1007/s11150-022-09601-},
  url={https://ideas.repec.org/a/kap/reveho/v20y2022i3d10.1007_s11150-022-09601-1.html}
}

@TECHREPORT{harley,
title = {Who Telecommutes? Where is the Time Saved Spent?},
author = {Frazis, Harley},
year = {2020},
institution = {Bureau of Labor Statistics},
type = {Economic Working Papers},
number = {523},
url = {https://EconPapers.repec.org/RePEc:bls:wpaper:523}
}

@Article{restrepo,
  author={Brandon J. Restrepo and Eliana Zeballos},
  title={{The effect of working from home on major time allocations with a focus on food-related activities}},
  journal={Review of Economics of the Household},
  year=2020,
  volume={18},
  number={4},
  pages={1165-1187},
  month={December},
  doi={10.1007/s11150-020-09497-},
  url={https://ideas.repec.org/a/kap/reveho/v18y2020i4d10.1007_s11150-020-09497-9.html}
}

@inproceedings{wu,
  title={The Impact of Work from Home (WFH) on Workload and Productivity in Terms of Different Tasks and Occupations},
  author={Hongyue Wu and Yunfeng Chen},
  booktitle={Interacci{\'o}n},
  year={2020},
  url={https://api.semanticscholar.org/CorpusID:221980403}
}

@article {hackney,
	Title = {Working in the digital economy: A systematic review of the impact of work from home arrangements on personal and organizational performance and productivity},
	Author = {Hackney, Amy and Yung, Marcus and Somasundram, Kumara G and Nowrouzi-Kia, Behdin and Oakman, Jodi and Yazdani, Amin},
	DOI = {10.1371/journal.pone.0274728},
	Number = {10},
	Volume = {17},
	Year = {2022},
	Journal = {PloS One},
	ISSN = {1932-6203},
	Pages = {e0274728},
	URL = {https://europepmc.org/articles/PMC9555618},
}

@techreport{skynova,
  author = {Skynova, Inc.},
  title = {Work-from-home lunch habits},
  year = 2021,
  url = {https://www.skynova.com/blog/work-from-home-lunch-habits},
  urldate = {2021-08-11}
}

@techreport{aksoy,
 title = "Working from Home Around the World",
 author = "Aksoy, Cevat Giray and Barrero, Jose Maria and Bloom, Nicholas and Davis, Steven J and Dolls, Mathias and Zarate, Pablo",
 institution = "National Bureau of Economic Research",
 type = "Working Paper",
 series = "Working Paper Series",
 number = "30446",
 year = "2022",
 month = "September",
 doi = {10.3386/w30446},
 URL = "http://www.nber.org/papers/w30446",
}

@article{bloom,
Author = {Bloom, Nicholas and Davis, Steven J. and Zhestkova, Yulia},
Title = {COVID-19 Shifted Patent Applications toward Technologies That Support Working from Home},
Journal = {AEA Papers and Proceedings},
Volume = {111},
Year = {2021},
Month = {May},
Pages = {263-66},
DOI = {10.1257/pandp.20211057},
URL = {https://www.aeaweb.org/articles?id=10.1257/pandp.20211057}}

\clearpage

\appendix
\counterwithin{figure}{section}
\counterwithin{table}{section}

\begin{figure}[!htb]
\section{Appendix}
	\begin{center}
		\includegraphics[width=\textwidth]{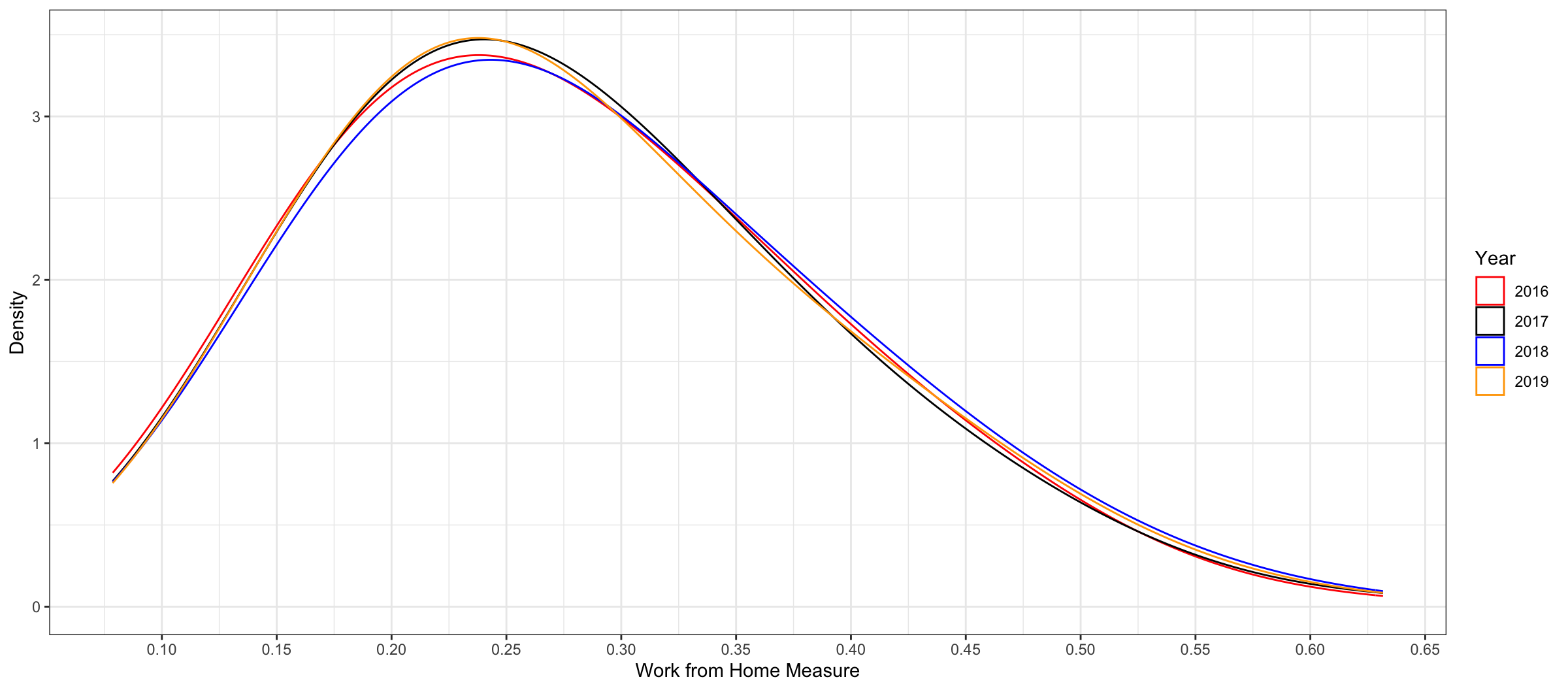} 
	\end{center}
        \caption{Distribution of WFH measure across district by year}\label{fig:wfh2016}
\end{figure}

\begin{longtable}{cp{10cm}c}
\caption{WFH measure by occupation (2-digit)}\label{tab:wfh2}\\
\toprule
KSCO & Occupation & WFH (in \%) \\
\midrule
11 & Senior Public Officials and Senior Corporate Officials &  $0$\\
12 & Public, Business Administration, Marketing Management Occupations & $100$ \\
13 & Professional Services Management Occupations & $88.95$ \\
14 & Construction, Electricity and Production Related Managers & $42.45$ \\
15 & Sales and Customer Service Managers & $76.35$ \\
21 & Science Professionals and Related Occupations & $17.9$ \\
22 & Information and Communication Professionals and Technical Occupations & $99.2$ \\
23 & Engineering Professionals and Technical Occupations & $19.77$ \\
24 & Health, Social Welfare and Religion Related Occupations & $23.47$ \\
25 & Education Professionals and Related Occupations & $99.72$ \\
26 & Legal and Administrative Occupations & $100$ \\
27 & Business and Finance Professionals and Related Occupations & $98.23$ \\
28 & Culture, Arts and Sports Professionals and Related Occupations & $63.75$ \\
31 & Administration and Accounting Related Occupations & $94.49$ \\
32 & Financial Clerical Occupations & $100$ \\
33 & Legal and Inspection Occupations & $100$ \\
39 & Customer Service, Information Desk, Statistical Survey and Other Clerical Occupations & $90.03$ \\
41 & Police, Fire Fighting and Security Related Service Occupations & $46.45$ \\
42 & Caregiving, Health and Personal Service Workers & $13.13$ \\
43 & Transport and Leisure Services Occupations & $0$ \\
44 & Cooking and Food Service Occupations & $33.34$ \\
51 & Sales Occupations & $63.56$ \\
52 & Store Sales and Rental Sales Occupations & $12.35$ \\
53 & Mobile, Door to Door and Street Sales Related Occupations & $71.75$ \\
61 & Agricultural, Livestock Related Skilled Occupations & $0$ \\
62 & Skilled Forestry Occupations & $0$ \\
63 & Skilled Fishery Occupations & $0$ \\
71 & Food Processing Related Trades Occupations & $0$ \\
72 & Textile, Clothing and Leather Related Trade Occupations & $0.60$ \\
73 & Wood and Furniture, Musical Instrument and Signboard Related Trade Occupations & $0$ \\
74 & Metal Coremakers Related Trade Occupations & $0$ \\
75 & Transport and Machine Related Trade Occupations & $0$ \\
76 & Electric and Electronic Related Trade Occupations & $0$ \\
77 & Information and Communications Technology Related Occupations & $0$ \\
78 & Construction and Mining Related Trade Occupations & $0$ \\
79 & Other Technical Occupations & $0$ \\
81 & Food Processing Related Machine Operating Occupations & $0$ \\
82 & Textile and Shoe Related Machine Operating Occupations & $0$ \\
83 & Chemical Related Machine Operating Occupations & $0$ \\
84 & Metal and Nonmetal Related Machine Operating Occupations & $0$ \\
85 & Machine Production and Related Machine Operating Occupation & $0$ \\
86 & Electrical and Electronic Related Machine Operating Occupations & $0$ \\
87 & Driving and Transport Related Occupations & $41.90$ \\
88 & Water Treatment and Recycling Related Operating Occupation & $0$ \\
89 & Wood, Printing and Other Machine Operating Occupations & $0.67$ \\
91 & Construction and Mining Related Elementary Occupations & $0$ \\
92 & Transport Related Elementary Occupations & $0$ \\
93 & Production Related Elementary Occupations & $0$ \\
94 & Cleaning and Guard Related Elementary Occupations & $0$ \\
95 & Household Helpers, Cooking Attendants and Sales Related Elementary Workers & $0$ \\
96 & Agriculture, Forestry, Fishery and Other Service Elementary Occupations & $21.44$ \\
\bottomrule
\end{longtable}

\begin{table}[H]
\caption{WFH measure by occupation (1-digit) from \cite{wfhkor}}\label{tab:wfh}
\centering
\begin{tabular}{clc}
\toprule
    KSCO & Occupation & WFH (in \%) \\
\midrule
    1 & Managers &  $65.45$\\
    2 & Professionals & $68.76$ \\
    3 & Clerical Support Workers & $100.0$ \\
    4 & Services Workers & $1.33$ \\
    5 & Sales Workers & $36.15$ \\
    6 & Skilled Agricultural, Forestry and Fishery Workers & $0$ \\
    7 & Craft and Related Trades Workers & $0$ \\
    8 & Plant and Machine Operators and Assemblers & $0$ \\
    9 & Elementary Occupations & 0 \\
\bottomrule \\
\end{tabular}
\end{table}

\end{document}